\documentclass[pra,prl,twocolumn,superscriptaddress,floatfix,showpacs,10pt,letterpaper]{revtex4}
\usepackage{graphicx} \usepackage{subfigure} \usepackage{amssymb}
\usepackage{verbatim} \usepackage{amsmath} \usepackage{amscd}
\usepackage{amssymb} 

\begin{document}

\title{Local unitary versus local Clifford equivalence of stabilizer
and graph states}

\author{Bei Zeng}
\affiliation{Department of Physics, Massachusetts Institute of
Technology, Cambridge, MA 02139, USA}

\author{Hyeyoun Chung}
\affiliation{Department of Electrical Engineering, Massachusetts
Institute of Technology, Cambridge, MA 02139, USA}

\author{Andrew W. Cross}
\affiliation{Department of Electrical Engineering, Massachusetts
Institute of Technology, Cambridge, MA 02139, USA} \affiliation{IBM
Research Division, T. J. Watson Research Center, P. O. Box 218,
Yorktown Heights, NY 10598}

\author{Isaac L. Chuang}
\affiliation{Department of Physics, Massachusetts Institute of
Technology, Cambridge, MA 02139, USA} \affiliation{Department of
Electrical Engineering, Massachusetts Institute of Technology,
Cambridge, MA 02139, USA}

\date{\today}

\begin{abstract} The equivalence of stabilizer states under local transformations is of fundamental interest in understanding properties and uses of entanglement. Two stabilizer states are equivalent under the usual stochastic local operations and classical communication criterion if and only if they are equivalent under local unitary (LU) operations. More surprisingly, under certain conditions, two LU equivalent stabilizer states are also equivalent under local Clifford (LC) operations, as was shown by Van den Nest et al. [Phys. Rev. \textbf{A71}, 062323]. Here, we broaden the class of stabilizer states for which LU equivalence implies LC equivalence ($LU\Leftrightarrow LC$) to include all stabilizer states represented by graphs with neither cycles of length $3$ nor $4$. To compare our result with Van den Nest et al.'s, we show that any stabilizer state of distance $\delta=2$ is beyond their criterion. We then further prove that $LU\Leftrightarrow LC$ holds for a more general class of stabilizer states of $\delta=2$. We also explicitly construct graphs representing $\delta>2$ stabilizer states which are beyond their criterion: we identify all $58$ graphs with up to $11$ vertices and construct graphs with $2^m-1$ ($m\geq 4$) vertices using quantum error correcting codes which have non-Clifford transversal gates. \end{abstract}

\pacs{03.67.Pp, 03.67.Mn, 03.67.Lx} \maketitle 


\section{Introduction}

Quantum entanglement, a phenomenon that has no counterpart in the
classical realm, is widely recognized as an important resource in
quantum computing and quantum information theory\cite{Nielsen}. {\it
Stabilizer states} form a particularly interesting class of
multipartite entangled states, which play important roles in areas
as diverse as quantum error correction\cite{Gottesman},
measurement-based quantum computing, and cryptographic
protocols\cite{Hans1,Hans2,Hans3,Hans4}. A stabilizer state on $n$
qubits is defined as the common eigenstate of its {\it stabilizer}:
a maximally abelian subgroup of the $n$-qubit Pauli group
$\mathcal{P}_n$ generated by the tensor products of the Pauli
matrices and the identity\cite{Gottesman}. Recently, a special
subset of stabilizer states (known as {\it graph states} due to
their association with mathematical graphs) has become the subject
of intensive study, and has proven to be useful in several fields of
quantum information theory\cite{Hans4,Werner}.

Despite their importance in quantum information science,
multipartite entangled states are still far from being well
understood\cite{Nielsen}. The study of multipartite entanglement has
usually focused on determining the equivalence classes of entangled
states under local operations, but there are too many such
equivalence classes under local unitary (LU) operations for a direct
classification to be practical. The most commonly studied set of
local operations are the invertible stochastic local operations
assisted with classical communication (SLOCC), which yield a much
smaller number of equivalence classes. For example, for three
qubits, there are only two classes of fully entangled states under
SLOCC, while $5$ real parameters are needed to specify the
equivalence classes under LU operations\cite{Vidal,Acin}. However,
the number of parameters needed to specify the equivalence classes
under SLOCC grows exponentially with $n$, where $n$ is the number of
qubits, so that specifying the equivalence classes for all states
rapidly becomes impractical for $n \geq 4$\cite{Verstra}.

For stabilizer states, a more tractable set of operations to study
is the local Clifford (LC) group, which consists of the local
unitary operations that map the Pauli group to itself under
conjugation. In addition to forming a smaller class of operations,
the local Clifford group has the additional advantage that the
transformation of stabilizer states under LC operations can be
reduced to linear algebra in a binary framework, which greatly
simplifies all the necessary computations\cite{Hans4}.

It has been conjectured that any two stabilizer states which are LU
equivalent are also LC equivalent (i.e. $LU \Leftrightarrow LC$
holds for every stabilizer state). If this were true, all of the
advantages of working with the local Clifford group would be
preserved when studying equivalences under an arbitrary local
unitary operation. Due to its far-reaching consequences, proving
that the $LU \Leftrightarrow LC$ equivalence holds for all
stabilizer states is possibly one of the most important open
problems in quantum information theory.

Graph states may prove to play a pivotal role in the proof of this
conjecture, as it has been shown that every stabilizer state is LC
equivalent to some graph state\cite{Moor1}. Therefore, if it could
be shown that $LU \Leftrightarrow LC$ holds for all graph states, it
would follow that $LU \Leftrightarrow LC$ holds for all stabilizer
states as well. Furthermore, it has been shown that an LC operation
acting on a graph state can be realized as a simple local
transformation of the corresponding graph, and that the orbits of
graphs under such local transformations can be calculated
efficiently\cite{Moor1,DANIELSEN,Moor2}. These results indicate that
if the $LU \Leftrightarrow LC$ equivalence holds for all graph
states, any questions concerning stabilizer states could be restated
in purely graph theoretic terms. This would make it possible to use
tools from graph theory and combinatorics to study the entanglement
properties of stabilizer states, and to tackle problems which may
have been too difficult to solve using more traditional approaches.

An important step towards a proof has been taken by Van den Nest et
al.\cite{Moor3}, who have shown that two LU equivalent stabilizer
states are also equivalent under LC operations if they satisfy a
certain condition, known as the Minimal Support Condition (MSC),
which ensures that their stabilizers possess some sufficiently rich
structure. They also conjecture that states which do not satisfy the
MSC will be rare, and therefore difficult to find.

\begin{figure}[htbp]
\includegraphics[width=3.00in]{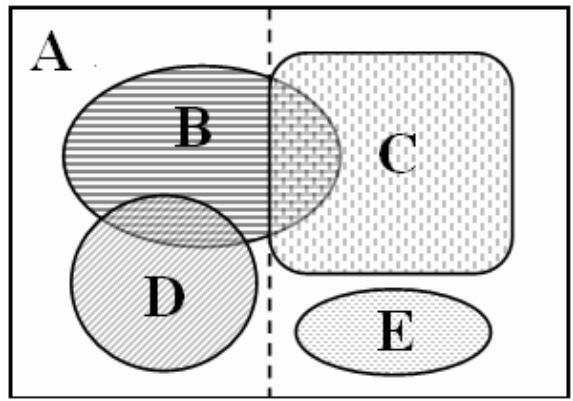}
\caption{Relations between theorems presented in this paper. A: all
graph states (there is a dashed line in the middle of A: the area
left of the line are graphs of distance $\delta=2$ and the right
area of the line are $\delta>2$ graphs); B: $LU\Leftrightarrow LC$
graphs given by Main Theorem; C: $LU\Leftrightarrow LC$ graphs given
by Van den Nest et al.'s criterion; D: $LU\Leftrightarrow LC$ graphs
of $\delta=2$ given by Theorem 2; E: Examples of $\delta>2$ graphs
beyond the MSC, given in Sec. IV, whose $LU\Leftrightarrow LC$
equivalence remains open.} \label{paperdiag}
\end{figure}

In this paper, we seek to make some progress towards a proof of the
$LU \Leftrightarrow LC$ conjecture, by proving that the $LU
\Leftrightarrow LC$ equivalence holds for all stabilizer states
whose corresponding graphs contain neither cycles of length $3$ nor
$4$. We also give some results complementary to those of Van den
Nest et al., by proving that the MSC does not hold for stabilizer
states of distance $\delta=2$, and by explicitly constructing states
of distance $\delta > 2$ which also fail to satisfy the MSC. Our
classification of stabilizer states is summarized in
Fig.~\ref{paperdiag}, which illustrates the relationship between the
subsets covered by our results and those of Van den Nest et al., as
well as those states for which the $LU \Leftrightarrow LC$
equivalence remains open.

Our paper is organized as follows: we first present some background
information on graph states and stabilizers in
Sec.~\ref{sec:prelim}. In Sec.~\ref{sec:maintheorem} we prove our
{\bf Main Theorem}, which states that $LU \Leftrightarrow LC$ holds
for any graph state (and hence, any stabilizer state) whose
corresponding graph contains neither cycles of length $3$ nor $4$.
We go on to prove that all stabilizer states with distance
$\delta=2$ fail to satisfy the MSC, whereas all stabilizer states
with $\delta > 2$ which satisfy the hypotheses of our {\bf Main
Theorem} do satisfy the MSC. We conclude Sec.~\ref{sec:maintheorem}
by using the proof of our {\bf Main Theorem} to show that $LU
\Leftrightarrow LC$ still holds for a particular subset of
stabilizer states with $\delta=2$. In Sec.~\ref{sec:dg2}, we provide
explicit examples of stabilizer states with distance $\delta > 2$
which fail to satisfy the MSC: we identify all $58$ graphs of up to
$11$ vertices which do not meet this condition, and construct two
other series of graphs beyond the MSC for $n = 2^m - 1$ $(m \geq 4)$
from quantum error correcting codes with non-Clifford transversal
gates. We conclude in Sec.~\ref{sec:concl}.

\section{Preliminaries} \label{sec:prelim}

Before presenting our \textbf{Main Theorem}, we state some
preliminaries in this section. We discuss the stabilizer formalism
and graph states in Sec. IIA. Then in Sec. IIB we introduce the
concept of minimal supports and Van den Nest et al.'s criterion.


\subsection{Stabilizers states and graph states}

The $n$-qubit Pauli group ${\mathcal P}_n$ consists of all $4\times
4^n$ local operators of the form $R= \alpha_R R_1\otimes\dots\otimes
R_n$, where $\alpha_R\in\{\pm 1, \pm i\}$ is an overall phase factor
and $R_i$ is either the $2\times 2$ identity matrix $I$ or one of
the Pauli matrices $X$, $Y$, or $Z$. We can write $R$ as $\alpha_R
(R_1)_1 (R_2)_2\dots (R_n)_n$ or $\alpha_R R_1 R_2\dots R_n$ when it
is clear what the qubit labels are. The $n$-qubit Clifford group
${\mathcal L}_n$ is the group of $n\times n$ unitary matrices that
map ${\mathcal P}_n$ to itself under conjugation.

A stabilizer ${\mathcal S}$ in the Pauli group ${\mathcal P}_n$ is
defined as an abelian subgroup of ${\mathcal P}_n$ which does not
contain $-I$. A stabilizer consists of $2^m$ Hermitian Pauli
operators for some $m\leq n$. As the operators in a stabilizer
commute with each other, they can be diagonalized simultaneously and
moreover, if $m=n$, then there exists a unique state $|\psi\rangle$
on $n$ qubits such that $R|\psi\rangle = |\psi\rangle$ for every
$R\in{\mathcal S}$. Such a state $|\psi\rangle$ is called the
stabilizer state and the group ${\mathcal S}={\mathcal
S}(|\psi\rangle)$ is called the stabilizer of $|\psi\rangle$. A
stabilizer state can also be viewed as a self-dual code over $GF(4)$
under the trace inner product\cite{DANIELSEN}. The distance $\delta$
of the state is the weight of the minimum weight element in
$\mathcal{S}(|\psi\rangle)$\cite{DANIELSEN}.

Two $n$-qubit states $|\psi\rangle$ and $|\psi'\rangle$ are said to
be local unitary (LU) equivalent if there exists an LU operation
\begin{equation}
\mathcal{U}_n=\bigotimes\limits_{i=1}^n U_i\label{LUO}
\end{equation}
which maps $|\psi'\rangle$ to $|\psi\rangle$.

Two $n$-qubit states $|\psi\rangle$ and $|\psi'\rangle$ are said to
be local Clifford (LC) equivalent if there exists an LU operation in
the Clifford group

\begin{equation}
\mathcal{K}_n=\bigotimes\limits_{i=1}^n K_i\label{LCO}
\end{equation}
which maps $|\psi'\rangle$ to $|\psi\rangle$, where $K_i\in\mathcal{L}_1$ for $i=1, \ldots,n$.

Throughout the paper we will use $\mathcal{U}_n$ and $\mathcal{K}_n$
to denote operations of the form Eq. (\ref{LUO}) and (\ref{LCO}),
respectively.

Graph states are a special kind of stabilizer state associated with
graphs\cite{Hans4}. A graph $G$ consists of two types of elements,
namely vertices ($V$) and edges ($E$). Every edge has two endpoints
in the set of vertices, and is said to connect or join the two
endpoints. The degree of a vertex is the number of edges ending at
that vertex. A path in a graph is a sequence of vertices such that
from each vertex in the sequence there is an edge to the next vertex
in the sequence. A cycle is a path such that the start vertex and
end vertex are the same. The length of a cycle is the number of
edges that the cycle has.

For every graph $G$ with $n$ vertices, there are $n$ operators
$R_a^G\in \mathcal{P}_n$ for $a=1,2,\ldots,n$ defined by
\begin{equation} R_a^G= X_{a} \bigotimes_{\{a,b\} \in E}
Z_{b},\label{Graph} \end{equation}

It is straightforward to show that any two $R_a^G$s commute, hence
the group generated by $\{R_a^G\}_{a=1}^n$ is a stabilizer group
$\mathcal{S}$ and stabilizes a unique state $|\psi_G\rangle$. We
call each $R_a^G$ the standard generator associated with vertex $a$
of graph $G$. Throughout the paper we use $|\psi_G\rangle$ to denote
the unique state corresponding to a given graph $G$.

Any stabilizer state is local Clifford (LC) equivalent to some graph
states\cite{Moor1}. Thus, it suffices to prove $LU\Leftrightarrow
LC$ for all graph states in order to show that $LU\Leftrightarrow
LC$ for all stabilizer states.


\subsection{Minimal supports}

The support $supp(R)$ of an element $R\in {\mathcal
S}(|\psi\rangle)$ is the set of all $i\in \{1,...,n\}$ such that
$R_i$ differs from the identity. Let $\omega=\{i_1, \dots, i_k\}$ be
a subset of $\{1, \dots, n\}$. Tracing out all qubits of
$|\psi\rangle$ outside $\omega$ gives the mixed state
\begin{equation}
\rho_{\omega}(\psi) = \frac{1}{2^{|\omega|}}\sum_{R\in {\mathcal
S}(|\psi\rangle), supp(R)\subseteq\omega}R \label{eq:msfour} \,.
\end{equation}

Using the notation $U_{\omega}=U_{i_1}\otimes\ldots\otimes U_{i_k}$,
it follows from $\mathcal{U}_n|\psi'\rangle=|\psi\rangle$ that
\begin{equation}
U_{\omega}\rho_{\omega}(\psi')U_{\omega}^{\dagger}=\rho_{\omega}(\psi)\,.
\label{eq:msfive}
\end{equation}

A minimal support of ${\mathcal S}(|\psi\rangle)$ is a set
$\omega\subseteq\{1,...,n\}$ such that there exists an element in
${\mathcal S}(|\psi\rangle)$ with support $\omega$, but there exist
no elements with support strictly contained in $\omega$. An element
in ${\mathcal S}(|\psi\rangle)$ with minimal support is called a
minimal element. We denote by $A_{\omega}(|\psi\rangle)$ the number
of elements $R\in{\mathcal S}(|\psi\rangle)$ with $supp(R)=\omega$.
Note that $A_{\omega}(|\psi\rangle)$ is invariant under LU
operations\cite{Moor3}. We use $\mathcal{M}(|\psi\rangle)$ to denote
the subgroup of ${\mathcal S}(|\psi\rangle)$ generated by all the
minimal elements. The following \textbf{Lemma 1} is given in
\cite{Moor3}.

[\textbf{Lemma 1}]: Let $|\psi\rangle$ be a stabilizer state and let
$\omega$ be a minimal support of ${\mathcal S}(|\psi\rangle)$. Then
$A_{\omega}(|\psi\rangle)$ is equal to 1 or 3 and the latter case
can only occur if $|\omega|$ is even.

If $\omega$ is a minimal support of ${\mathcal S}(|\psi\rangle)$, it
follows from the proof of \textbf{Lemma 1} in \cite{Moor3} that the
minimal elements with support $\omega$, up to an LC operation, must
have one of the following two forms:
\begin{eqnarray}
A_{\omega}(|\psi\rangle)=1\ &:&\ Z^{\otimes\omega}\nonumber\\
A_{\omega}(|\psi\rangle)=3\ &:&\ \{X^{\otimes\omega},
(-1)^{(|\omega|/2)}Y^{\otimes\omega}, Z^{\otimes\omega}\} \,.
\label{ME}
\end{eqnarray}

Eqs.(\ref{eq:msfour}), (\ref{eq:msfive}) and (\ref{ME}) directly
lead to the following \textbf{Fact 1}, which was originally proved
by Rains in \cite{Rains}:

[\textbf{Fact 1}]: If $|\psi'\rangle$ and $|\psi\rangle$ are LU
equivalent stabilizer states, i.e.
$\mathcal{U}_n|\psi'\rangle=|\psi\rangle$, then for each minimal
support $\omega$, the equivalence $\mathcal{U}_n$ must take the
group generated by all the minimal elements of support $\omega$ in
$\mathcal{S}(|\psi'\rangle)$ to the corresponding group generated by
all the minimal elements of support $\omega$ in
$\mathcal{S}(|\psi\rangle)$.

Based on the above \textbf{Fact 1}, the following \textbf{Theorem 1}
was proven in \cite{Moor3} as their main result:

[\textbf{Theorem 1}]: Let $|\psi\rangle$ be a fully entangled
stabilizer state for which all three Pauli matrices $X,Y,Z$ occur on
every qubit in $\mathcal{M}(|\psi\rangle)$. Then every stabilizer
state $|\psi'\rangle$ which is LU equivalent to $|\psi\rangle$ must
also be LC equivalent to $|\psi\rangle$.

The condition given in \textbf{Theorem 1}, that all three Pauli
matrices $X,Y,Z$ occur on every qubit in
$\mathcal{M}(|\psi\rangle)$, is called the minimal support condition
(MSC).

For any LU operation $\mathcal{U}_n=\bigotimes\limits_{i=1}^n U_i$
which maps another stabilizer state $|\psi'\rangle$ to the
stabilizer state $|\psi\rangle$, the proof of \textbf{Theorem 1}
further specifies the following

[\textbf{Fact 2}]: If all three Pauli matrices $X,Y,Z$ occur on the
$j$th qubit in $\mathcal{M}(|\psi\rangle)$, then $U_j$ must be a
Clifford operation. Therefore, if the MSC condition is satisfied for
$|\psi\rangle$, then $\mathcal{U}_n$ must be an LC operation.

In \cite{Moor3} it is also shown that although $n$-GHZ
states\cite{GHZ} do not possess this structure, $LU\Leftrightarrow
LC$ still holds.

\section{The main theorem}
\label{sec:maintheorem}

We now present the new criterion we have found for the
$LU\Leftrightarrow LC$ equivalence of graph states. Sec. IIIA, IIIB,
IIIC, and IIID are devoted to proving the main result of the paper.
An algorithm for constructing the LC operation
$\mathcal{K}_n=\bigotimes\limits_{i=1}^n K_i$, where
$K_i\in\mathcal{L}_1$ for any $i$, is given in Sec. IIIE and
\textbf{Theorem 2}, which covers additional $LU\Leftrightarrow LC$
equivalences for $\delta=2$ graphs beyond the main theorem, is given
in Sec. IIIF.

The main result of the paper is the following:

[\textbf{Main Theorem}]: $LU\Leftrightarrow LC$ equivalence holds
for any graph $G$ with neither cycles of length 3 nor 4.

[\textbf{Proof}]: In order to prove that $LU\Leftrightarrow LC$
holds for $|\psi_G\rangle$, we will show that for any stabilizer
state $|{\psi}_G'\rangle$ satisfying
$\mathcal{U}_n|{\psi}_G'\rangle=|{\psi}_G\rangle$, there exists an
LC operation $\mathcal{K}_n$ such that
$\mathcal{K}_n|{\psi}_G'\rangle=|{\psi}_G\rangle$. The proof is
presented in several sections below, ending in Sec. IIID on page 9.

We prove this theorem constructively, i.e. we construct
$\mathcal{K}_n$ explicitly from the given $\mathcal{U}_n$,
$|\psi_G\rangle$, and $|\psi'_G\rangle$. Before giving the details
of our proof, we give a brief outline of our strategy. We will
assume that throughout our proof that all graphs have neither cycles
of length $3$ nor $4$.

First, we show that any graph of distance $\delta>2$ satisfies the
MSC, hence $LU\Leftrightarrow LC$ holds for them. However, we will
also show that any graph of distance $\delta=2$ is beyond the MSC.
Therefore, we only need to prove the \textbf{Main Theorem} for
$\delta=2$ graphs.

We then partition the vertex set $V(G)$ of graph $G$ into subsets
$\{V_1(G),V_2(G),V_3(G),V_4(G)\}$ as defined later. We show that for
all vertices $v\in V_3(G)\cup V_4(G)$, the operator $U_v$ in
$\mathcal{U}_n$ must be a Clifford operation, i.e.
$U_v\in\mathcal{L}_1$. For vertices $v\in V_1(G)\cup V_2(G)$, we
will give a procedure, called the standard procedure, for
constructing $K_v$. In effect, this corresponds to an ``encoding" of
any vertex $v\in V_2$ and all the degree one vertices $w\in V_1$ to
which $v$ is connected into a repetition code (i.e. ``deleting" the
degree one vertices from $G$), and then a ``decoding" of the code.

\begin{figure}[htbp]
\includegraphics[width=2.00in]{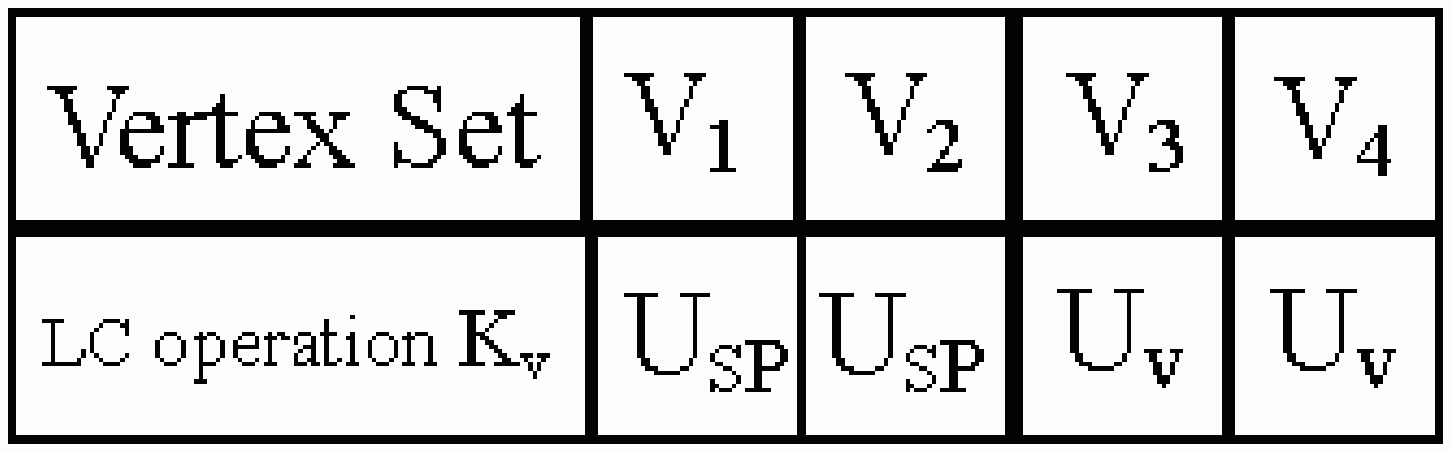}
\caption{An illustration of the construction of $\mathcal{K}_n$:
we will simply choose $K_v=U_v$ for all $v\in V_3\cup V_4$,
and use the standard procedure(SP) to construct $K_v=U_{SP}$ for all $v\in V_1\cup V_2$.}
\label{proofdiag}
\end{figure}

We illustrate the proof idea in Fig.~\ref{proofdiag}. Due to some
technical reasons, we first show $U_v\in\mathcal{L}_1$ for all $v\in
V_4$ in Sec. IIIA. Then we give the standard procedure in Sec. IIIB.
We use an example to show explicitly how the procedure works, with
explanations of why this procedure actually works in general.
Finally, in Sec. IIIC we show that $U_v\in\mathcal{L}_1$ for all
$v\in V_3(G)\cup V_4(G)$, and construct $K_v$ for all $v\in
V_1(G)\cup V_2(G)$ from the standard procedure.

\begin{figure}[htbp]
\includegraphics[width=3.00in]{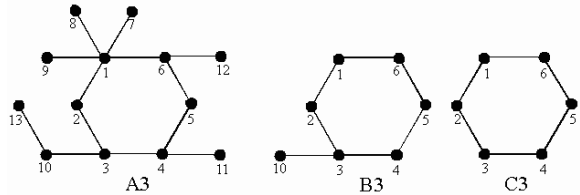}
\caption{Examples of the partition: $V_1(A3)=\{7,8,9,11,12,13\}$,
$V_2(A3)=\{1,4,6,10\}$, $V_3(A3)=\{5\}$ and $V_4(A3)=\{2,3\}$;
$V_1(B3)=\{10\}$, $V_2(B3)=\{3\}$, $V_3(B3)=\emptyset$ and
$V_4(B3)=\{1,2,4,5,6\}$; $C3$ is a graph of $\delta=3$ hence
$V_1(C3)=V_2(C3)=V_3(C3)=\emptyset$, and
$V_4(C3)=V(C3)=\{1,2,3,4,5,6\}$.}
\label{thepartition}
\end{figure}

The four types of vertices we use for a graph $G$ are defined as
follows. $V_1(G)$ is the degree one vertices of $G$. $V_2(G)$ is the
set of vertices $V_2(G)=\{v|v$ connects to some $w\in V_1(G)\}$. The
set $V_3(G)$ is given by $V_3(G)=\{v|v$ not in $V_1(G)$, and $v$
only connects to $w\in V_2(G)\}$. Finally, the set $V_4(G)$ is
defined by $V_4(G)=V(G)\setminus(V_1(G)\cup V_2(G) \cup V_3(G))$.
For convenience, we also apply this partitioning of vertices to
$\delta>2$ graphs, hence $V(G)=V_4(G)$. Fig.~\ref{thepartition}
gives an example of such partitions.

\subsection{$\delta>2$ and $\delta=2$ graphs and Case $V_4$}

We first provide some lemmas which lead to a proof of the
\textbf{Main Theorem} for $\delta>2$ graphs. Then we show that all
$\delta=2$ graphs are beyond the MSC.

\subsubsection{$\delta>2$ graphs}

[\textbf{Lemma 2}]: For a vertex $v\in V(G)$ which is unconnected to
any degree one vertex, if it is neither in cycles of length $3$ nor
$4$, and then $R_v$ is the only minimal element of support
$supp(R_v)$.

[\textbf{Proof}]: Suppose the vertex $v$ connects to vertices
$i_1,i_2,\cdots i_k,$, then $R_v=X_vZ_{i_1}Z_{i_2}\cdots Z_{i_k}$.
If there exists an element $S_m\in\mathcal{S}(|\psi_G\rangle)$ such
that $supp(S_m)\subseteq supp(R_v)$, then $S_m$ must be expressed as
a product of elements in $\{R_v,R_{i_1},R_{i_2},\cdots R_{i_k}\}$.
However since $v$ is neither in any cycle of length $3$ nor $4$,
then any product of elements in $\{R_v,R_{i_1},R_{i_2},\cdots
R_{i_k}\}$ (except $R_v$ itself) must contain at least one Pauli
operator $\alpha_j$ acting on the $j$th qubit where $j$ is not in
$supp(R_v)$. $\square$

This directly leads to the following \textbf{Lemma 3} for $\delta>2$
graphs:

[\textbf{Lemma 3}]: For any graph $G$ with $\delta>2$, if there are
neither cycles of length $3$ nor $4$, then $G$ satisfies the MSC,
and hence $LU\Leftrightarrow LC$ holds for $G$.

[\textbf{Proof}]: Since $\delta>2$, then all vertices $v\in V(G)$
are unconnected to any degree one vertices. Then by \textbf{Lemma
2}, $\mathcal{M}(|\psi\rangle)=\mathcal{S}(|\psi\rangle)$, and
therefore the MSC is satisfied. $\square$

\textbf{Lemma 2} tells us that for any vertex $v\in V_4(G)$, we must
have $U_v\in\mathcal{L}_1$, according to \textbf{Fact 2}.
\textbf{Lemma 3} shows that we only need to prove the \textbf{Main
Theorem} for graphs of $\delta=2$.

\subsubsection{$\delta=2$ graphs}

[\textbf{Proposition 1}]: Stabilizer states with distance $\delta=2$
are beyond the MSC.

[\textbf{Proof}]: A stabilizer state $|\psi\rangle$ with $\delta =
2$ has at least one weight two element in its stabilizer
$\mathcal{S}(|\psi\rangle)$. We denote one such weight two element
by $\alpha_i \beta_k$, where $\alpha_j$ and $\beta_k$ are one of the
three Pauli operators $X,Y,Z$ on the $j$th and $k$th qubits
respectively, up to an overall phase factor of $\pm 1$ or $\pm i$.
Now consider any element $R$ in $\mathcal{S}(|\psi \rangle)$ with a
support $\omega$ such that $\omega \cap \{j,k\} \neq \emptyset$. We
can write $R$ in the form $R_1R_2\cdots R_n$ where each $R_i$ is
either the identity matrix $I$ or one of the Pauli matrices $X,Y,Z$,
up to an overall phase factor of $\pm 1$ or $\pm i$. Then there are
three possibilities: (i) If $\omega \cap \{j,k\}$ is $\{j\}$ or
$\{k\}$, then since $R$ commutes with $\alpha_j \beta_k$, the
operator $R_j$ ($R_k$) can only be $\alpha_j$ ($\beta_k$), up to an
overall phase factor of $\pm 1$ or $\pm i$. (ii) If $\omega =
\{j,k\}$, then since $R$ commutes with $\alpha_j \beta_k$, we either
have $R_jR_k = \alpha_j' \beta_k'$, where $\alpha_j'$ anticommutes
with $\alpha_j$ and $\beta_k'$ anticommutes with $\beta_k$, or
$R_jR_k = \alpha_j \beta_k$. The former is impossible, as the whole
graph is connected, so the latter must hold. (iii) If $\omega$
strictly contains $\{j,k\}$, then $R$ is not a minimal element. It
follows that in $\mathcal{M}(|\psi\rangle)$, only $\alpha_j$ appears
on the $j$th qubit and only $\beta_k$ appears on the $k$th qubit,
showing that $\mathcal{S}(|\psi\rangle)$ is beyond the MSC.$\square$

Furthermore, the local unitary operation $\mathcal{U}_n$ which maps
another $\delta=2$ stabilizer state $|\psi'\rangle$ to
$|\psi\rangle$ is not necessarily in the Clifford group,
particularly on the $j$th and $k$th qubits. Note that it is always
true for any angle $\theta$ that \begin{equation}
\alpha_j(\theta)\beta_k(-\theta)|\psi\rangle=e^{i\alpha_j\theta}e^{-i\beta_k\theta}|\psi\rangle=|\psi\rangle.\label{free}
\end{equation}

To interpret \textbf{Proposition 1} in view of graphs, it is noted
that any fully connected graph $G$ with degree one vertices
represents a graph state $|\psi_G\rangle$ of $\delta=2$. Therefore,
a graph with degree one vertices is beyond the MSC. In particular,
for a graph $G$ with neither cycles of length $3$ nor $4$, each
weight two element in $\mathcal{S}(|\psi_G\rangle)$ corresponds to
the standard generator of a degree one vertex in $G$.


\subsection{Case $V_1\cup V_2$: The standard procedure}

The main idea behind the standard procedure is to convert the
$LU$-equivalent stabilizer states $|\psi_G\rangle$ and
$|\psi'_G\rangle$ into the corresponding (LC equivalent) canonical
forms for which we can prove $LU\Leftrightarrow LC$ by applying
``encoding" and ``decoding" methods. We can then work backwards from
those canonical forms to prove that $LU\Leftrightarrow LC$ for
$|\psi_G\rangle$.

\begin{figure}[htbp] \includegraphics[width=2.50in]{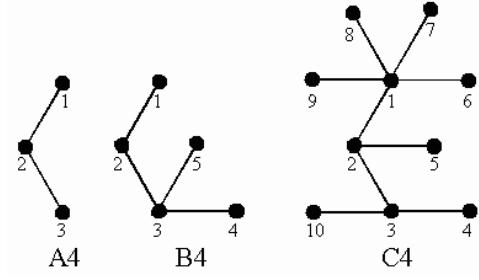} \caption{$A4$ is a subgraph of both $B4$ and $C4$.} \label{simpleGHZ} \end{figure}

We use a simple example, as shown in graph B4 of
Fig.~\ref{simpleGHZ}, to demonstrate how the standard procedure
works. The standard procedure decomposes into five steps. In each
step, we also explain how the step works for the general case.

Note that $|\psi_{A4}\rangle$ is a GHZ state; hence
$LU\Leftrightarrow LC$ holds. The standard generator of the
stabilizer for graph $A4$ is $\{XZI,ZXZ,IZX\}$. However, as we will
see later in step 4, $LU\Leftrightarrow LC$ for $A4$ does not
guarantee that $LU\Leftrightarrow LC$ for $B4$.

We now prove $LU\Leftrightarrow LC$ for $|\psi_{B4}\rangle$.

\subsubsection{Step 1: Transform into a new basis by LC operation}

It is straightforward to show \begin{equation}
|\psi_{B4}\rangle=\frac{1}{2^{5/2}}\sum\limits_{a_j=0,1}(-1)^{f(E)}|a_1a_2a_3a_4a_5\rangle,
\end{equation} where $f(E)=a_1a_2+a_2a_3+a_3a_4+a_3a_5$, which is
determined by the the edge set $E(B4)$.

Performing Hadamard transform on the fourth and fifth qubits, we get
\begin{equation} |\tilde{\psi}_{B4}\rangle=H_4\otimes
H_5|\psi_B\rangle=\frac{1}{\sqrt{2}}(|\xi_{0}\rangle|000\rangle+|\xi_{1}\rangle|111\rangle)\label{newform},
\end{equation} where \begin{eqnarray}
|\xi_0\rangle&=&\frac{1}{2}(|00\rangle+|01\rangle+|10\rangle-|11\rangle)\nonumber\\
|\xi_1\rangle&=&\frac{1}{2}(|00\rangle-|01\rangle+|10\rangle+|11\rangle).
\end{eqnarray}

The form of $|\tilde{\psi}_{B4}\rangle$ in Eq.(\ref{newform}) is not
hard to understand. By performing $H_4\otimes H_5$, the standard
generator of $|\psi_{B4}\rangle$ will be transformed to
$\{Z_3Z_4,Z_3Z_5...\}$, hence only the terms of $|000\rangle$ and
$|111\rangle$ appear on the qubits $3,4,5$. Furthermore, for the
supports $\omega_1=(3,4),\omega_2=(3,5)$, we have
$A_{\omega_1}(|\psi_{B4}\rangle)=A_{\omega_2}(|\psi_{B4}\rangle)=1$.

For any other stabilizer state which is LU equivalent to
$|\psi_{B4}\rangle$, there exist an LU operation $\mathcal{U}_5$
such that $\mathcal{U}_5|\psi_{B4}'\rangle=|\psi_{B4}\rangle$.
According to \textbf{Fact 1}, for the supports
$\omega_1=(3,4),\omega_2=(3,5)$, there must also be
$A_{\omega_1}(|\psi_{B4}'\rangle)=A_{\omega_2}(|\psi_{B4}'\rangle)=1$.
Suppose the corresponding minimal elements of $\omega_1,\omega_2$
are $\alpha_3\beta_4,\alpha_3\gamma_5$ respectively, then there
exist $F_3,F_4,F_5\in\mathcal{L}_1$, such that
$(F_3\alpha_3F_3^{\dagger})\otimes (F_4\beta_4F_4^{\dagger})=Z_3
Z_4$, $(F_3\alpha_3F_3^{\dagger})\otimes
(F_5\gamma_5F_5^{\dagger})=Z_3Z_5$. Therefore, we have
\begin{eqnarray} |\tilde{\psi}_B'\rangle&=&F_3\otimes F_4\otimes
F_5|\psi'_B\rangle\nonumber\\
&=&\frac{1}{\sqrt{2}}(|\chi_{0}\rangle|000\rangle+|\chi_{1}\rangle|111\rangle)\label{newform2},
\end{eqnarray} where $|\chi_0\rangle$ and $|\chi_1\rangle$ are two
states of qubits $1$ and $2$.

The states $|\tilde{\psi}_B\rangle$ and $|\tilde{\psi}_B'\rangle$
given in Eqs.(\ref{newform},\ref{newform2}) are then called
canonical forms of $|{\psi}_B\rangle$ and $|{\psi}_B'\rangle$,
respectively.

Then we have \begin{equation}
\tilde{\mathcal{U}}_5|\tilde{\psi}_B'\rangle=|\tilde{\psi}_B\rangle,\label{tildeU}
\end{equation} where \begin{equation}
\tilde{\mathcal{U}}_5=H_4\otimes H_5\mathcal{U}_5
F_3^{\dagger}\otimes F_4^{\dagger}\otimes
F_5^{\dagger}\label{tildeUO} \end{equation} i.e. $\tilde{U}_1=U_1$,
$\tilde{U}_2=U_2$, $\tilde{U}_3=U_3F_3^{\dagger}$,
$\tilde{U}_4=H_4U_4F_4^{\dagger}$,
$\tilde{U}_5=H_5U_5F_5^{\dagger}$.

Eq.(\ref{tildeU}) is then our new starting point, since
$|\psi'_{B4}\rangle$ and $|\psi_{B4}\rangle$ are LC equivalent if
and only if $|\tilde{\psi}'_{B4}\rangle$ and
$|\tilde{\psi}_B\rangle$ are LC equivalent, then we can always get
the former when we prove the latter by reversing Eq.
(\ref{tildeUO}), as we will do from eqs. (\ref{tildeK}) to
(\ref{finalK}).

Note the procedure of getting Eq.(\ref{tildeU}) is general, i.e. we
can always do the same thing for any $\delta=2$ graph state and its
LU equivalent graph states. To be more precise, for a general graph
$G$ of $n$ vertices, consider a vertex $a \in V_2(G)$, and let
$N(a)$ be the set of all degree one vertices in $V(G)$ which connect
to $a$. If the size of this set is $|N(a)| = k$, then without loss
of generality we can rename the qubits so that the vertices $a$ and
$b \in N(a)$ are represented by the last $k+1$ qubits of
$|\psi_G\rangle$.

Applying the Hadamard transform $\tilde{H}_a = \bigotimes_{b \in
N(a)} H_b$ to $|\psi_G\rangle$ gives a new stabilizer state
$|\tilde{\psi}_G^{(a)}\rangle$ as shown below. \begin{eqnarray}
\tilde{H}_a|\psi_G\rangle&=&|\tilde{\psi}_G^{(a)}\rangle\nonumber\\
&=&\frac{1}{\sqrt{2}}(|\xi_0\rangle|0\rangle^{\bigotimes (k+1)} +
|\xi_1\rangle|1\rangle^{\bigotimes (k+1)}), \label{eq:equation8}
\end{eqnarray} where $|\xi_0\rangle$ and $|\xi_1\rangle$ are
two states of the other $n-(k+1)$ qubits.

Similarly, for any stabilizer state $|\psi'_G\rangle$ which is LU
equivalent to $|\psi_G\rangle$, i.e.
$\mathcal{U}_n|\psi'_G\rangle=|\psi_G\rangle$, there must exist
$F_a,F_b\in\mathcal{L}_1$ (for all $b\in N(a)$) such that
\begin{equation} (F_a\alpha_a F_a^{\dagger})\otimes (F_b\beta_b
F_b^{\dagger}) = Z_aZ_b, \label{eq:equation10} \end{equation} for
$\alpha_a\beta_b\in\mathcal{S}(|\psi'_G\rangle)$.

Define $\tilde{F}_a = F_a \bigotimes_{b \in N(a)} F_b$, we have
\begin{eqnarray}
\tilde{F}_a|\psi'_G\rangle&=&|\tilde{\psi'}_G^{(a)}\rangle\nonumber\\
&=&\frac{1}{\sqrt{2}}(|\chi_0\rangle|0\rangle^{\bigotimes (k+1)} +
|\chi_1\rangle|1\rangle^{\bigotimes (k+1)}), \label{eq:equation9}
\end{eqnarray} where $|\chi_0\rangle$ and $|\chi_1\rangle$ are
two states of the other $n-(k+1)$ qubits.

We apply the above procedure for all $a\in V_2(G)$. Define
$\tilde{H}=\bigotimes_{a\in V_2(G)}{\tilde{H}_a}$ and
$\tilde{F}=\bigotimes_{a\in V_2(G)}{\tilde{F}_a}$, we get
\begin{eqnarray}
\tilde{H}|\psi_G\rangle&=&|\tilde{\psi}_G\rangle\nonumber\\
\tilde{F}|\psi_G\rangle&=&|\tilde{\psi}_G'\rangle, \label{tildepsi}
\end{eqnarray}

Now define \begin{equation} \tilde{\mathcal{U}}_n =
\bigotimes_{i=1}^n \tilde{U}_i, \label{eq:equation12} \end{equation}
where $\tilde{U}_i = U_i$ for all $i \in V_3(G)\cup V_4(G)$,
$\tilde{U}_a = U_aF_a^{\dagger}$ for all $a\in V_2(G)$, and
$\tilde{U}_b = H_bU_bF_b^{\dagger}$ for all $b \in N(a)$. We then
have $\tilde{\mathcal{U}}_n|\tilde{\psi}_G'\rangle =
|\tilde{\psi}_G\rangle$.

It can be seen that $|\psi_G'\rangle$ and $|\psi_G\rangle$ are LC
equivalent if and only if $|\tilde{\psi}_G'\rangle$ and
$|\tilde{\psi}_G\rangle$ are LC equivalent. Therefore, we can use
the states $|\tilde{\psi}_G'\rangle$ and $|\tilde{\psi}_G\rangle$ as
our new starting point.

Our current situation is summarized in the following diagram. \[
\begin{CD} |\psi_G\rangle @<\mathcal{U}_n = \bigotimes_{i=1}^n U_i<<
|\psi_G'\rangle\\ @V\tilde{H} = \bigotimes_{a \in V_2(G)}
\tilde{H}_aVV @VV\tilde{F} = \bigotimes_{a \in V_2(G)}
\tilde{F}_aV\\ |\tilde{\psi}_G\rangle
@<\tilde{\mathcal{U}}_n=\bigotimes_{i=1}^n \tilde{U}_i<<
|\tilde{\psi}_G'\rangle\\ \end{CD}\]

\subsubsection{Step 2: Encode into repetition codes}

Now we can encode the qubits $3,4,5$ into a single logical qubit,
i.e. $|0_L\rangle=|000\rangle$, $|1_L\rangle=|111\rangle$. Define
$|\bar{\psi}_{B4}\rangle=(|\xi_{0}\rangle|0_L\rangle+|\xi_{1}\rangle|1_L\rangle)$,
and
$|\bar{\psi}_{B4}'\rangle=(|\chi_{0}\rangle|0_L\rangle\rangle+|\chi_{1}\rangle|1_L\rangle)$,
then both $|\bar{\psi}_{B4}\rangle$ and $|\bar{\psi}_{B4}'\rangle$
are $3$-qubit stabilizer states. Especially,
$|\bar{\psi}_{B4}\rangle$ is exactly the graph state
$|\psi_{A4}\rangle$ represented by graph $A4$. Now Eq.(\ref{tildeU})
becomes \begin{equation}
\bar{\mathcal{U}_3}|\bar{\psi}_{B4}'\rangle=|\bar{\psi}_{B4}\rangle,\label{tildeUL}
\end{equation} where $\bar{\mathcal{U}_3}=U_1\otimes U_2\otimes
U_L^{(3)}$, and $U_L^{(3)}$ is a logical operation acting on the
logical qubit, which must be of some special forms as we discuss
below. The upper index $(3)$ indicates that we may understand this
logical qubit $L$ as being the $3$rd qubit in graph $A4$.

Due to \textbf{Fact 1}, we must have \begin{eqnarray}
\tilde{U}_3Z_3\tilde{U}_3^{\dagger}\otimes\tilde{U}_4Z_4\tilde{U}_4^{\dagger}&=&Z_3Z_4\nonumber\\
\tilde{U}_3Z_3\tilde{U}_3^{\dagger}\otimes\tilde{U}_5Z_5\tilde{U}_5^{\dagger}&=&Z_3Z_5\label{fact}
\end{eqnarray} which means either \begin{eqnarray}
\tilde{U}_3Z_3\tilde{U}_3^{\dagger}&=&Z_3\nonumber\\
\tilde{U}_4Z_4\tilde{U}_4^{\dagger}&=&Z_4\nonumber\\
\tilde{U}_5Z_5\tilde{U}_5^{\dagger}&=&Z_5,\label{plus}
\end{eqnarray} which gives
$\tilde{U}_3=diag(1,e^{i\theta_1}),\tilde{U}_4=diag(1,e^{i\theta_2}),\tilde{U}_5=diag(1,e^{i\theta_3})$
for some $\theta_1,\theta_2,\theta_3$, or \begin{eqnarray}
\tilde{U}_3Z_3\tilde{U}_3^{\dagger}&=&-Z_3\nonumber\\
\tilde{U}_4Z_4\tilde{U}_4^{\dagger}&=&-Z_4\nonumber\\
\tilde{U}_5Z_5\tilde{U}_5^{\dagger}&=&-Z_5\label{minus}
\end{eqnarray} which gives $\tilde{U}_3=diag(1,e^{i\theta_1})X_3$,
$\tilde{U}_4=diag(1,e^{i\theta_2})X_4$,
$\tilde{U}_5=diag(1,e^{i\theta_3})X_5$ for some
$\theta_1,\theta_2,\theta_3$.

Therefore, we must have
$U_L^{(3)}=diag(1,e^{i(\theta_1+\theta_2+\theta_3)})$ if
Eq.(\ref{plus}) holds, or
$U_L^{(3)}=diag(1,e^{i(\theta_1+\theta_2+\theta_3)})X_L^{(3)}$ if
Eq.(\ref{minus}) holds.

Note the procedure of getting Eq.(\ref{tildeUL}) and the result of
the possible forms that $U_L$ possesses is also general. Recall that
we have two states of the form given in Eq.~(\ref{eq:equation8}) and
Eq.~(\ref{eq:equation9}), we can encode the qubits $a$ and $b \in
N(a)$ into a single logical qubit, by writing $|0_L\rangle =
|0\rangle^{\otimes(k+1)}$ and $|1_L\rangle =
|1\rangle^{\otimes(k+1)}$. We can then define two new stabilizer
states $|\bar{\psi}_G^{(a)}\rangle$ and
$|\bar{\psi}_G'^{(a)}\rangle$, given by \begin{eqnarray}
|\bar{\psi}_G^{(a)}\rangle &=& |\xi_0\rangle|0_L\rangle +
|\xi_1\rangle|1_L\rangle,\nonumber\\ |\bar{\psi}_G'^{(a)}\rangle &=&
|\chi_0\rangle|0_L\rangle + |\chi_1\rangle|1_L\rangle.
\label{equation14} \end{eqnarray} Both are stabilizer states of $m$
qubits, where $m = n-k$. In particular, $|\bar{\psi}_G^{(a)}\rangle$
is represented by a graph which is obtained by deleting all the
vertices $b\in N(a)$ from $G$.

We can see that $|\bar{\psi}_G'^{(a)}\rangle$ and
$|\bar{\psi}_G^{(a)}\rangle$ are related by \begin{eqnarray}
\bar{\mathcal{U}}_m^{(a)}|\bar{\psi}_G'^{(a)}\rangle =
|\bar{\psi}_G^{(a)}\rangle,\label{equation15} \end{eqnarray} where
$\bar{\mathcal{U}}_m^{(a)}=\bigotimes_{i=1}^{m-1} {U}_i\otimes
U_L^{(a)}$, and $U_L^{(a)}$ is a logical operation acting on the
logical qubit $a$.

Similarly, we can place some restrictions on the form taken by
$U_L^{(a)}$. By {\bf Fact 1}, we have
\begin{equation}
\tilde{U}_aZ_a\tilde{U}_a^{\dagger} \otimes
\tilde{U}_bZ_b\tilde{U}_b^{\dagger} = Z_aZ_b \label{eq:equation16}
\end{equation}
for all $b \in N(a)$. This means that either
\begin{eqnarray}
\tilde{U}_a &=& diag(1, e^{i\theta_a}),\nonumber\\
\tilde{U}_b &=& diag(1, e^{i\theta_b}) \label{equation19}
\end{eqnarray} for all $b \in N(a)$ and some $\theta_a, \theta_b$, which gives
\begin{equation}
U_L^{(a)} = diag(1,e^{i\theta}),\label{eq:equation20}
\end{equation}
where $\theta = \theta_a + \sum_{b \in N(a)} \theta_b$, or
\begin{eqnarray}
\tilde{U}_a &=& diag(1, e^{i\theta_a})X_a,\nonumber\\
\tilde{U}_b &=& diag(1, e^{i\theta_b})X_b \label{equation19a}
\end{eqnarray} for all $b \in N(a)$ and some $\theta_a, \theta_b$, which gives
\begin{equation}
U_L^{(a)} = diag(1,e^{i\theta})X_L^{(a)},\label{eq:equation22}
\end{equation} where $\theta = \theta_a + \sum_{b \in N(a)} \theta_b$.

Now again we apply the above encoding procedure for all $a\in
V_2(G)$. This leads to two $m$-qubit stabilizer states
$|\bar{\psi}_G\rangle$ and $|\bar{\psi}_G'\rangle$, where
$m=n-|V_1(G)|$. In particular, $|\bar{\psi}_G^{(a)}\rangle$ is
represented by a graph which is obtained by deleting all the degree
one vertices from $G$. Define \begin{equation}
\bar{\mathcal{U}}_m=\bigotimes_{i=1}^{m-|V_2(G)|}
{U}_i\bigotimes_{a\in V_2(G)} U_L^{(a)}, \end{equation} we then have
\begin{eqnarray} \bar{\mathcal{U}}_m|\bar{\psi}_G'\rangle =
|\bar{\psi}_G\rangle,\label{Umbarg} \end{eqnarray}

After this step of our standard procedure, our situation is as shown
below: \[ \begin{CD} |\psi_G\rangle @<\mathcal{U}_n =
\bigotimes_{i=1}^n U_i<< |\psi_G'\rangle\\ @V\tilde{H} =
\bigotimes_{a \in V_2(G)} \tilde{H}_aVV @VV\tilde{F} =\bigotimes_{a
\in V_2(G)}\tilde{F}_bV\\ |\tilde{\psi}_G\rangle
@<\tilde{\mathcal{U}}_n=\bigotimes_{i=1}^n \tilde{U}_i<<
|\tilde{\psi}_G'\rangle\\ @V encode VV @VV encode V \\
|\bar{\psi}_G\rangle @<\bar{\mathcal{U}}_m<< |\bar{\psi}_G'\rangle
\end{CD}\]

\subsubsection{Step 3: Show that $U_L\in \mathcal{L}_1$}

We then further show that $U_L^{(3)}\in \mathcal{L}_1$, which means
$\theta_1+\theta_2+\theta_3=0,\pi/2,\pi,3\pi/2$. Consider the
minimal element $Z_2X^{(3)}_L$, it is the standard generator of
graph $A4$ associated with the (logical) qubit $3$. Then we have
$A_{\omega=(2,3)}=1$ holds for both $|\bar{\psi}_{B4}\rangle$ and
$|\bar{\psi}_{B4}'\rangle$. Furthermore, $Z_2X^{(3)}_L$ is the only
minimal element of $\omega=supp(Z_2X^{(3)}_L)=(2,3)$ according to
\textbf{Proposition 1}. If $U^{(3)}_L$ is not in $\mathcal{L}_1$,
then $U^{(3)}_LR^{(3)}_LU^{(3)\dagger}_L\neq X^{(3)}_L$ for any
$R^{(3)}_L\in\mathcal{P}_1$, which contradicts \textbf{Fact 1}. It
is not hard to see that the fact of $U_L\in \mathcal{L}_1$ is also
general.

We now show $U_L^{(3)}\in \mathcal{L}_1$ can also be induced by
local Clifford operations on the qubits $3,4,5$. This can be simply
given by $diag(1,e^{i(\theta_1+\theta_2+\theta_3)})_3\otimes
I_4\otimes I_5$ if Eq.(\ref{plus}) holds, or
$diag(1,e^{i(\theta_1+\theta_2+\theta_3)})_3X_3\otimes X_4\otimes
X_5$ if Eq.(\ref{minus}) holds.

In the general case, it is shown in \textbf{Lemma 2} that for a
graph with neither cycles of length $3$ nor $4$, the standard
generator $R_v$ of any vertex $v$ which is unconnected to degree one
vertices will be the only minimal element of $supp(R_v)$. Then due
to the form of $U_L^{(a)}$ in Eq.(\ref{eq:equation22}), we conclude
that for a general graph with neither cycles of length $3$ nor $4$,
any induced $U^{(a)}_L$ must be in $\mathcal{L}_1$. Similarly, each
$U_L^{(a)}\in \mathcal{L}_1$ can also be induced by local Clifford
operations on the qubits $\{\{a\}\cup b\in N(a)\}$. This can be
simply given by $diag(1,e^{i\theta})_a\bigotimes_{b\in N(a)} I_b$ if
Eq.(\ref{eq:equation20}) holds, or
$diag(1,e^{i\theta})_aX_a\bigotimes_{b\in N(a)}X_b$ if
Eq.(\ref{eq:equation22}) holds.

\subsubsection{Step 4: Construct a logical LC operation relating $|\bar{\psi}_G\rangle$ and $|\bar{\psi}'_G\rangle$}

In this step, we start from the general case first and then go back
to our example of the graph $A4$.

For a general graph $G$, of which $V_3(G)$ and $V_4(G)$ are not both
empty sets, we show that for $|\bar{\psi}_G\rangle$, $U_i$ must be
in $\mathcal{L}_1$ for any $i$ which is not a logical operation. To
see this, note we have already shown in Sec. III A,
$U_v\in\mathcal{L}_1$ for all $v\in V_4(G)$. And we are going to
show in Sec. III C that $U_v\in\mathcal{L}_1$ for all $v\in V_3(G)$.
We also have applied step 1 and 2 to each $a\in V_2(G)$ to obtain
$U_L^{(a)}$. As shown in step 3, $U_L^{(a)}\in\mathcal{L}_1$, hence
we have $\bar{\mathcal{U}}_m=\bigotimes_{i=1}^{m-|V_2(G)|}
{U}_i\bigotimes_{a\in V_2(G)} U_L^{(a)}$ is an LC operation such
that
$\bar{\mathcal{U}}_m|\bar{\psi}'_G\rangle=|\bar{\psi}_G\rangle$.

Now we go back to our example. Note for graph $A4$, we have already
shown that $U_L^{(3)}$ is a Clifford operation. If we could further
show that $U_1$ and $U_2$ are also Clifford operations, then
$\bar{\mathcal{U}}_3=U_1\otimes U_2\otimes U_L^{(3)}$ is an LC
operation which maps $|\bar{\psi}_{B4}'\rangle$ to
$|\bar{\psi}_{B4}\rangle$.

However, for graph $B4$, $V_3(B4)=V_4(B4)=\emptyset$, i.e. the
vertices $1$ and $2$ are neither in $V_3(B4)$ nor $V_4(B4)$. Then we
have to show that although $U_1$ and $U_2$ themselves do not
necessarily be Clifford operations, there do exist
$\tilde{K}_1,\tilde{K}_2\in \mathcal{L}_1$, such that
\begin{equation} \tilde{K}_1\otimes \tilde{K}_2\otimes
U_L^{(3)}|\bar{\psi}_{B4}'\rangle=|\bar{\psi}_{B4}\rangle.\label{GHZLULC}
\end{equation}

This can be checked straightforwardly due to the simply form of
$|\bar{\psi}_{B4}\rangle=\frac{1}{\sqrt{2}}(|0_x00_x\rangle+|1_x11_x\rangle)$,
where $|0_x (1_x)\rangle=\frac{1}{\sqrt{2}}(|0\rangle\pm|1\rangle)$.
And we know $|\bar{\psi}_{B4}'\rangle$ is also a $3$-qubit GHZ
state, hence $U_1$ and $U_2$ can only be of very restricted forms.
To be more concrete, for instance, for
$|\bar{\psi}_{B4}'\rangle=\frac{1}{\sqrt{2}}(|000_y\rangle+|111_y\rangle)$,
where $|0_y(1_y)\rangle=\frac{1}{\sqrt{2}}(|0\rangle\pm
i|1\rangle)$, there could be $U_1=H_1 diag(1,e^{-i\theta})_1$,
$U_2=diag(1,e^{i\theta})_2$ and $U_L^{(3)}=diag(1,-i)_3$, i.e.
\begin{eqnarray} H_1 diag(1,e^{-i\theta})_1\otimes
diag(1,e^{i\theta})_2\otimes diag(1,-i)_3\nonumber\\
\times\frac{1}{\sqrt{2}}(|000_y\rangle+|111_y\rangle)
=\frac{1}{\sqrt{2}}(|0_x00_x\rangle+|1_x11_x\rangle). \end{eqnarray}

But we know \begin{eqnarray} H_1\otimes I_2&\otimes&
diag(1,-i)_3\nonumber\\
\times\frac{1}{\sqrt{2}}(|000_y\rangle&+&|111_y\rangle)\frac{1}{\sqrt{2}}(|0_x00_x\rangle+|1_x11_x\rangle).
\end{eqnarray}

Note other possibilities of $|\bar{\psi}_{B4}'\rangle$ (and the
possible corresponding $U_1$, $U_2$ and $U_L^{(3)}$) can also be
checked similarly.

One may ask why we do not also delete the vertex $1$ in graph $B4$
as we do in the general case, then it is likely that we are also
going to get a logical Clifford operation $U_L^{(2)}$ on the vertex
$2$. Then for the graph with only two vertices $2$ and $3$, we have
an LC operation $U_L^{(2)}\otimes U_L^{(3)}$. However, this is not
true due to the fact that the connected graph of only two qubits is
beyond our \textbf{Proposition 1}. Then in this case the argument in
step 3 no longer holds.

\subsubsection{Step 5: Decode $U_L^{(a)}$ to construct $\mathcal{K}_n$}

Finally, the following steps are natural and also general. We can
then choose $\tilde{K}_3=U^{(3)}_L$, and choose
$\tilde{K}_4=\tilde{K}_5=I$ if
$U^{(3)}_L=diag(1,e^{i(\theta_1+\theta_2+\theta_3)})$ or
$\tilde{K}_4=\tilde{K}_5=X$ if
$U^{(3)}_L=diag(1,e^{i(\theta_1+\theta_2+\theta_3)})X^{(3)}_L$,
which gives \begin{equation}
\tilde{\mathcal{K}}_5|{\tilde{\psi}_{B4}'}\rangle=|\tilde{\psi}_{B4}\rangle,\label{tildeK}
\end{equation} where
$\tilde{\mathcal{K}}_5=\bigotimes\limits_{i=1}^5 \tilde{K}_i$.

Define $\mathcal{K}_5=\bigotimes\limits_{i=1}^5 K_i$, where
$K_1=\tilde{K}_1$, $K_2=\tilde{K}_2$, $K_3=\tilde{K}_3F_3$,
$K_4=H_4\tilde{K}_4F_4$, $U_5=H_5\tilde{K}_5F_5$, then
\begin{equation} \mathcal{K}_5|{{\psi}_{B4}'}\rangle
=|{\psi}_{B4}\rangle,\label{finalK} \end{equation} which is desired.

In general, for each $a\in V_2(G)$ and all $b\in N(a)$, choose
$\tilde{K}_a=U^{(a)}_L$ and choose $\tilde{K}_b=I_b$ if
$U^{(a)}_L=diag(1,e^{i\theta})$, or $\tilde{K}_a=U^{(a)}_LX_a$ and
$\tilde{K}_b=X_b$ if $U^{(a)}_L=diag(1,e^{i\theta})X^{(a)}_L$.
Define \begin{equation} \tilde{\mathcal{K}}_n=\bigotimes_{i\in
V_3(G)\cup V_4(G)} U_i\bigotimes_{j\in V_1(G)\cup
V_2(G)}\tilde{K}_j, \end{equation} we have \begin{equation}
\tilde{\mathcal{K}}_n|\tilde{\psi}'_G\rangle=|\tilde{\psi}_G\rangle.
\end{equation}

Define $\mathcal{K}_n=\bigotimes\limits_{i=1}^n K_i$, where
$K_i=U_i$ for all $i\in V_2(G)\cup V_3(G)$; for each $a\in V_2(G)$,
$K_a=\tilde{K}_aF_a$ and $K_b=H_b\tilde{K}_bF_b$ for all $b\in
N(a)$,then \begin{equation}
\mathcal{K}_n|{\psi}'_G\rangle=|{\psi}_G\rangle, \end{equation}
which is desired.

Steps 3,4 and 5 are then summarized as the following diagram.

\[ \begin{CD} |\psi_G\rangle @<\mathcal{K}_n = \bigotimes_{i=1}^n K_i<< |\psi_G'\rangle\\ @A\tilde{H}^{\dagger} = \bigotimes_{a \in V_2(G)} \tilde{H}_a^{\dagger}AA @AA\tilde{F}^{\dagger} = \bigotimes_{a\in V_2(G)} \tilde{F}_a^{\dagger}A\\ |\tilde{\psi}_G\rangle @<\tilde{\mathcal{K}}_n=\bigotimes_{i=1}^n \tilde{K}_i<< |\tilde{\psi}_G'\rangle\\ @A decode AA @AA decode A \\ |\bar{\psi}_G\rangle @<\bar{\mathcal{U}}_m\in\mathcal{L}_1<< |\bar{\psi}_G'\rangle \end{CD}\]


\subsection{Case $V_3$}

Unlike the case that for $v\in V_4(G)$, where $U_v\in\mathcal{L}_1$
is guaranteed by \textbf{Lemma 2} and \textbf{Fact 2}, case $V_3$ is
more subtle. Note \textbf{Lemma 2} does apply for any $v\in V_3(G)$,
i.e. the standard generator $R_v$ is the only minimal element of
$supp(R_v)$, however for any $x\in N(v)$, $R_x$ is not in
$\mathcal{M}(|\psi\rangle)$ due to \textbf{Proposition 1}.

We now use the standard procedure to prove that
$U_v\in\mathcal{L}_1$ for all $v\in V_3$, thereby proving that
$LU\Leftrightarrow LC$ for $|\psi_G\rangle$. We use $\bar{G}$ to
denote the graph obtained by deleting all the degree one vertices
from $G$. Note for any $v\in V_3(G)$, there must be $v\in
V{(\bar{G})}$. Then there are three possible types of vertices in
$V_3$: type 1, $v\in V_2{(\bar{G})}$; type 2: $v\in V_4{(\bar{G})}$;
and type 3: $v\in V_3{(\bar{G})}$. We discuss all the three types in
Sec. 1, 2 and 3, respectively.

\subsubsection{Type 1}

\begin{figure}[htbp] \includegraphics[width=3.40in]{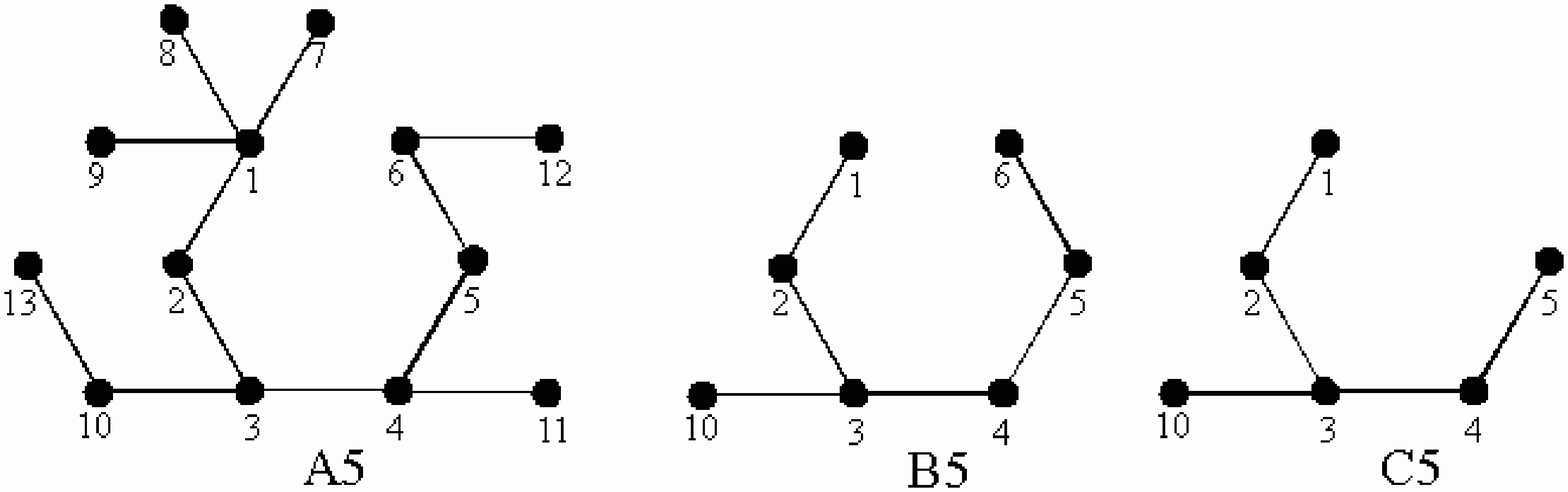} \caption{An example of type 1 vertices: for graph A5, $V_1(A5)=\{7,8,9,11,12,13\}$, $V_2(A5)=\{1,4,6,10\}$, $V_3(A5)=\{5\}$ which is type 1, and $V_4(A5)=\{2,3\}$.} \label{trees} \end{figure}

The subtlety of proving $v\in V_3$ for a type 1 vertex $v$ is that
we need to apply the standard procedure twice to make sure
$U_v\in\mathcal{L}_1$. We will demonstrate this with the following
example, to prove $LU\Leftrightarrow LC$ for graph $A5$ in
Fig.~\ref{trees}.

For $\mathcal{U}_{13}|\psi'_{A5}\rangle=|\psi_{A5}\rangle$, the
standard construction procedure will result in
$\bigotimes\limits_{i=1}^{6} V_i\otimes
V_{10}|\psi'_{B5}\rangle=|\psi_{B5}\rangle$, where
$V_i\in\mathcal{L}_1$ for $i=1,2,3,4,6,10$ and $V_5=U_5$. Now we
again use the construction procedure on qubit $5$ of $B_3$ and
encode the qubits $5,6$ into a single qubit $5$, as shown in Fig.
5C) ($C5$). This gives $\bigotimes\limits_{i=1}^{4} W_i\otimes
W_{5}\otimes W_{10}|\psi'_{C5}\rangle=|\psi_{C5}\rangle$, where
$W_i\in\mathcal{L}_1$ for $i=1,2,3,4,5,10$. Here $W_5$ is induced by
$V_5,V_6$ via a similar process as eqs. (12,13,14). Since
$V_6\in\mathcal{L}_1$, we must have $U_5=V_5\in\mathcal{L}_1$, as
desired.

In general we can prove $U_v\in\mathcal{L}_1$ for any type 1 vertex
$v\in V_3$ as we did for vertex $5$ in the above example of graph
$A5$. To be more precise, let $v\in V_3(G)$ be a vertex of type 1.
For each $v$, carrying out the standard procedure at all $x\in N(v)$
gives us a graph $G_1$. We know that each $U_L^{(x)}$ must be in
$\mathcal{L}_1$. Since $v\in V_2(\bar{G})$,
we then have a non-empty $N(v)\cap V_1(\bar{G})$. Again for $G_1$ we
carry out the standard procedure at $v$, giving us a graph $G_2$,
and each $U_L^{(v)}$ must be in $\mathcal{L}_1$. This gives
$U_v\in\mathcal{L}_1$ due to the form of $U_L^{(v)}$ in
eqs.(\ref{eq:equation20},\ref{eq:equation22}).

\subsubsection{Type 2}

Now we consider the type 2 vertices. We give an example first, to
prove that $LU\Leftrightarrow LC$ for graph $A3$ in Fig.
\ref{thepartition}. $A3$ is a graph without cycles of length 3 and
4, and represents a general graph with four types of vertices. $A3$
is very similar to $A5$, and has the same set of $V_1,V_2,V_3,V_4$
as $A5$. The only difference between the two graphs is that in $A3$,
vertices $1$ and $6$ are connected to each other. Therefore,
following the example for the graph $A5$ shows that for any
$\mathcal{U}_{13}|\psi'_{A3}\rangle=|\psi_{A3}\rangle$, the standard
construction procedure will result in $\bigotimes\limits_{i=1}^{6}
V_i\otimes V_{10}|\psi'_{B3}\rangle=|\psi_{B3}\rangle$, where
$V_i\in\mathcal{L}_1$ for $i=1,2,3,4,6,10$ and $V_5=U_5$. However,
from the structure of $B3$, it is easy to conclude that
$V_5=U_5\in\mathcal{L}_1$.

In general, we can prove $U_v\in\mathcal{L}_1$ for any type 2 vertex
$v\in V_3$ as we did for vertex $5$ in the above example of graph
$A3$. To be more precise, let $v\in V_3(G)$ be a vertex of type 2.
For each $v$, carrying out the standard procedure at all $x\in N(v)$
gives us a graph $G_1$. $G$ contains neither cycles of length $3$
nor $4$, so the same holds for $G_1$. Since $v\in V_4(\bar{G})$, we
have $v\in V_4(G_1)$. Due to \textbf{Lemma 2}, we conclude that
$U_v\in\mathcal{L}_1$.

\subsubsection{Type 3}

Now we consider the type 3 vertices. Let us first examine an
example. Consider the graph $A3'$ which is obtained by deleting
vertices $2$ and $13$ from graph $A3$. For this new graph with
$V(A3')=\{1,3,4,5,6,7,8,9,10,11,12\}$, we have
$V_1(A3')=\{7,8,9,10,11,12\}$, $V_2(A3')=\{1,3,4,6\}$,
$V_3(A3')=\{5\}$ and $V_4(A3')=\emptyset$. It is easy to see that
the vertex $5$ is of type 3. Carrying out the standard procedure at
vertices $4$ and $6$ gives a graph $A3''$, which is a subgraph of
$A3$ with $V(A3'')=\{1,3,4,5,6,7,8,9,10\}$. Now we see that $5\in
V_4(A3')$, and hence $U_5\in\mathcal{L}_1$ for any
$\bigotimes\limits_{i\in V(A3')} U_i$ which takes the graph state
$|\psi_{A3'}\rangle$ to another $11$-qubit stabilizer state.

In general, note that $v\in V_3(G)$ is of type 3 only when every
vertex $x\in N(v)$ not only connects to some degree one vertices,
but also connects to some vertices in $V_2(G)$. So the trick is to
perform the standard procedure only at all $x\in N(v)$. This gives a
graph $G_2$. Since $v\in V_3(\bar{G})$, we have $v\in V_4(G_2)$. Due
to our result in Sec. III A1, we conclude that
$U_v\in\mathcal{L}_1$.


\subsection{Some remarks}

To summarize, in general we first classify the vertices of $G$ into
four types ($V_1(G),V_2(G),V_3(G),V_4(G)$). To construct
$\mathcal{K}_n$, we choose $K_i=U_i$ for all $i\in V_3(G)\cup
V_4(G)$, and then apply the standard procedure to construct $K_i$
for all $i\in V_1(G)\cup V_2(G)$.

Note that for some graphs for which $V_3$ and $V_4$ are both empty
sets, for instance the graph $B4$ in Fig.{\ref{simpleGHZ}}, the
general procedure discussed in the above paragraph does not apply
directly. This special situation has already been discussed in
detail in Sec III B4.

This completes our proof of the \textbf{Main Theorem}.$\square$


\subsection{Algorithm for constructing $\mathcal{K}_n$}

The proof of our \textbf{Main Theorem} implies a constructive
procedure for obtaining the local Clifford operation $\mathcal{K}_n$
corresponding to a given local unitary operation $\mathcal{U}_n$.
This procedure is described below. For clarity, the operation
``$\times$ is used to denote standard matrix multiplication in
$SU(2)$.

[\textbf{Algorithm: Construction of} $\mathcal{K}_n$]:

\noindent \medskip{\bf CONSTRUCT-LC[$G$, $\mathcal{U}_n$]}:\\[2ex]

\noindent Input: A connected graph $G$ with no cycles of length 3 or 4; a stabilizer state $|\psi_G'\rangle$ and an LU operation $\mathcal{U}_n = \bigotimes_{i=1}^n U_i$ such that $\mathcal{U}_n |\psi_G'\rangle = |\psi_G\rangle$.\\[2ex]

\noindent Output: An LC operation $\mathcal{K}_n= \bigotimes_{i=1}^n K_i$ such that $\mathcal{K}_n |\psi_G'\rangle = |\psi_G\rangle$.\\[2ex]

\noindent 1. Partition $V(G)$ into subsets $V_1, V_2, V_3, V_4$.\\
2. Let $K_i \leftarrow U_i$ for all $i \in V_3 \cup V_4$.\\ 3. {\bf
for} each $v_2 \in V_2$:\\ 3.1 \hspace*{0.3cm}Calculate $B_{v_2} =
U_{v_2}^{\dagger}Z_{v_2}U_{v_2}$.\\ 3.2 \hspace*{0.3cm}Find any
$F_{v_2} \in \mathcal{L}_1$ such that
$F_{v_2}B_{v_2}F_{v_2}^{\dagger} = Z_{v_2}$.\\
3.3\hspace*{0.3cm}Calculate $\tilde{U}_{v_2} =
U_{v_2}F_{v_2}^{\dagger}$.\\ 3.4 \hspace*{0.3cm}Find
$\{w_1,\dotsc,w_k\} \subset V_1$ such that $\{w_j,v_2\} \in E(G)$\\
\hspace*{0.85cm}for all $1 \leq j \leq k$.\\ 3.5 \hspace*{0.3cm}{\bf
for} $j \leftarrow 1$ {\bf to} $k$\\ 3.5.1 \hspace*{0.55cm}Find any
$F_{w_j} \in \mathcal{L}_1$ such that
$F_{w_j}B_{w_j}F_{w_j}^{\dagger} = Z_{w_j}$.\\ 3.5.2 \hspace*{0.4cm}
Calculate $\tilde{U}_{w_j} = H_{w_j}U_{w_j}F_{w_j}^{\dagger}$.\\
3.5.3 \hspace*{0.3cm}{\bf end for}\\ 3.6 \hspace*{0.3cm}{\bf if}
$\tilde{U}_{v_2}$ is diagonal\\ 3.6.1 \hspace*{0.5cm}Calculate
$\tilde{K}_{v_2} =
\tilde{U}_{v_2}\times\tilde{U}_{w_1}\dotsc\times\tilde{U}_{w_k}$.\\
3.6.2 \hspace*{0.5cm}Let $\tilde{K}_{w_j} = I_{w_j}$ for all $j$.\\
3.6.3 \hspace*{0.5cm}Let $K_{v_2} = \tilde{K}_{v_2}F_{v_2}, K_{w_j}
= H_{w_j}\tilde{K}_{w_j}F_{w_j}$.\\ 3.7 \hspace*{0.3cm}{\bf else}\\
3.7.1 \hspace*{0.497cm}Calculate
$\tilde{K}_{v_2}=\tilde{U}_{v_2}X_{v_2}\times\tilde{U}_{w_1}X_{w_1}\dotsc\times\tilde{U}_{w_k}X_{w_k}$.\\
3.7.2 \hspace*{0.35cm} Let $\tilde{K}_{w_j} = X_{w_j}$ for all
$j$.\\ 3.7.3 \hspace*{0.45cm}Let $K_{v_2}=\tilde{K}_{v_2}F_{v_2},
K_{w_j} = H_{w_j}\tilde{K}_{w_j}F_{w_j}$.\\ 3.7.4
\hspace*{0.3cm}{\bf end if}\\ 3.8 {\bf end for} \\ 4. {\bf return}
$\mathcal{K}_n =\bigotimes_{i=1}^n K_i$.


\subsection{$\delta=2$ graphs beyond the main theorem}

In this section, we present a theorem regarding $LU\Leftrightarrow
LC$ for $\delta=2$ graphs. We again use $\bar{G}$ to denote the
graph obtained by deleting all the degree one vertices from $G$.

[\textbf{Theorem 2}]: $LU\Leftrightarrow LC$ holds for any
$\delta=2$ graph $G$ if $\bar{G}$ satisfies the MSC.

[\textbf{Proof}]: The proof is the same as the proof of the
\textbf{Main Theorem} in the special case where $V_3(G)$ is an empty
set. $\square$

Although the proof of \textbf{Theorem 2} is a special case of the
proof of the \textbf{Main Theorem}, \textbf{Theorem 2} is not a
corollary of the \textbf{Main Theorem}. It can be applied to many
$\delta=2$ graphs with cycles of length $3$ or $4$, since we know
that many $\delta>2$ graphs satisfy the MSC.

\section{$\delta>2$ graph states beyond the MSC} \label{sec:dg2}

From \textbf{Lemma 3}, we know that for graphs of $\delta>2$, our
\textbf{Main Theorem} is actually a corollary of \textbf{Theorem 1}.
Now an interesting question is: do there exist other graph states
with distance $\delta>2$ which are beyond the MSC? The answer is
affirmative. Below, in Sec. IVA, we present some examples for the
case $n\leq 11$ qubits. In Sec. IVB we construct two series of
$\delta>2$ graphs beyond the MSC for $n=2^m-1$ ($m\geq 4$) out from
error correcting codes with non-Clifford transversal gates. In Sec.
IVC, we briefly discuss the $LU\Leftrightarrow LC$ property for
$\delta>2$ graphs.


\subsection{$\delta>2$ graphs beyond the MSC for minimal $n$}

Generally the distance of a graph state can be upper bounded by $2
\left\lfloor \frac{n}{6} \right\rfloor + 1$ for a graph whose
elements in $\mathcal{S}$ have even weight, which only happens when
$n$ is even. For the other graphs, the distance is upper bounded by
$2 \left\lfloor \frac{n}{6} \right\rfloor + 1$, if $n \equiv 0 $ mod
6, $2 \left\lfloor \frac{n}{6} \right\rfloor + 3$, if $n \equiv 5 $
mod 6, and $2 \left\lfloor \frac{n}{6} \right\rfloor + 2$,
otherwise\cite{Rains2}.

\begin{figure}[htbp] \includegraphics[width=2.30in]{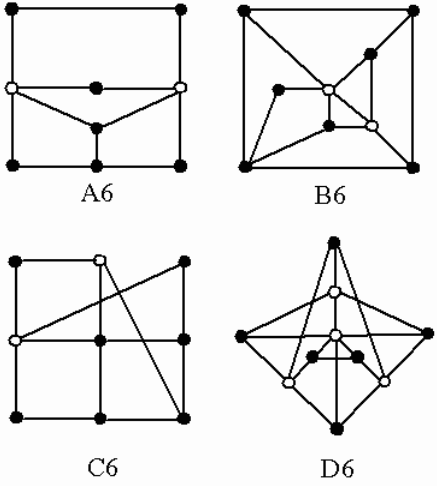} \caption{A6, B6, C6: Three $\delta=3$ graphs beyond the MSC for $n=9$; D6: The only one $\delta=4$ graph beyond the MSC for $n=10$. In each graph all the black vertices are minimal elements which are just generators of the corresponding $\mathcal{M}$, and all the white vertices are not in $\mathcal{M}$.} \label{distancegeq3} \end{figure}

Our numerical calculations show that there are no $\delta>2$ graphs
beyond the MSC for $n<9$. Among all the $440$ LC non-equivalent
connected graphs of $n=9$, there are only three $\delta>2$ graphs
beyond the MSC. All of them are of distance three, which are shown
as graphs A6, B6, and C6 in Fig.~\ref{distancegeq3}. Among all the
$3132$ LC non-equivalent connected graphs of $n=10$, there are only
nine $\delta>2$ graphs beyond the MSC. Eight of them are of distance
three, only one is of distance four. The distance four graph of
$n=10$ beyond the MSC is shown as graph D6 in
Fig.~\ref{distancegeq3}. Among all the $40457$ LC non-equivalent
connected graphs of $n=11$, there are only $46$ $\delta>2$ graphs
beyond the MSC. $37$ of them are of distance three and $9$ are of
distance four.

\subsection{Graphs derived from codes with non-Clifford transversal gates}

In this section we construct other two series of $\delta>2$ graphs
beyond the MSC for $n=2^m-1$ ($m\geq 4$) from error correcting codes
with non-Clifford transversal gates.

It is well-known that transversal gates on quantum codes, i.e.
logical unitary operations which could be realized via a bitwise
manner, is crucial for fault-tolerant quantum
computing\cite{Nielsen,Gottesman}. General single qubit transversal
gates on an $n$-qubit code $Q$ is of the form $\mathcal{U}_n$.
However, only Clifford transversal gates $\mathcal{K}_n$ are
relatively easy to find from symmetries of the
stabilizer\cite{Nielsen,Gottesman}, while it is hard to find
non-Clifford transversal gates for a given stabilizer code.

To construct the CSS code with transversal gates \begin{equation}
\exp\left(-i \frac{\pi}{2^{m-1}}\, {Z_L} \right) \cong
\bigotimes_{i=1}^{2^m-1} \exp\left(i \frac{\pi}{2^{m-1}}\,
Z_i\right), \end{equation} consider the first order punctured
Reed-Muller code $C_1=RM^* (1,m)$ with parameters
$[2^m-1,m+1,2^{m-1}-1]$ and its even subcode $C_2=even(RM^* (1,m))$
with parameters $[2^m-1,m,2^{m-1}]$\cite{MacWilliams}. It is
well-known that the dual code of $C_1$ is the binary Hamming code
with parameters $[2^{m}-1,2^m-1-m,3]$. Then this gives a series of
quantum codes with parameters $[2^m-1,1,3]$. For a given $m$, the
code is spanned by $|\bar{0}\rangle = \sum_{c\in C_2} |c\rangle$ and
$|\bar{1}\rangle = \sum_{c\in C_1-C_2} |c\rangle$. The computational
basis vectors on which $|\bar{0}\rangle$ has support have weight $0$
or $2^{m-1}$ and those of $|\bar{1}\rangle$ have weight $2^{m-1}-1$
or $2^m-1$\cite{Feng}. Therefore, $\exp{\left(-i \frac{\pi}{2^{m-1}}
{Z_L} \right)}$ is a valid transversal gate.

Similar to the classical Reed-Muller codes, from the point of view
of code parameters, these quantum codes become weaker as their
length increases. However, non-Clifford operations are not all
equal; some are more complex than others, even for fixed qubit
number. Note that $\exp\left(-i \frac{\pi}{2^{m-1}}\, {Z_L}
\right)\in \mathcal{C}_k$ with $k=m-1$, where $\mathcal{C}_k$ is
defined by \begin{equation} \mathcal{C}_{k+1}=\{U\in
U(\mathcal{H})|U\mathcal{C}_1 U^{\dagger}\subseteq \mathcal{C}_k\},
\end{equation} where $\mathcal{C}_1$ is the Pauli group, and
generally gates in $\mathcal{C}_k$ with larger $k$ are
stronger\cite{Got}. Hence it worths constructing codes with
transversal $\mathcal{C}_k$ gates for any $k$.

\begin{figure}[htbp] \includegraphics[width=3.00in]{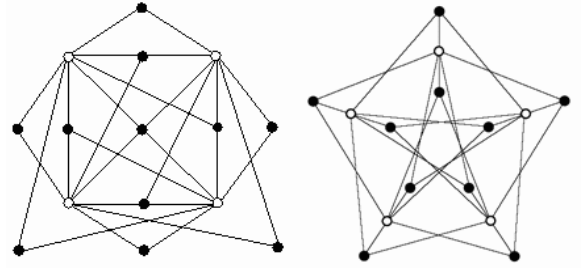} \caption{$d\geq 3$ graphs beyond the MSC. The left graph corresponds to the $|0_L\rangle$ state of the 15 qubit code with transversal $T$ gate. And the right one is a graph corresponds to $|+_L\rangle$, getting from \cite{Rau}. In each graph all the black vertices are minimal elements which are just generators of the corresponding $\mathcal{M}$, and all the white vertices are not in $\mathcal{M}$.} \label{15qubit} \end{figure}

\begin{figure}[htbp] \includegraphics[width=3.20in]{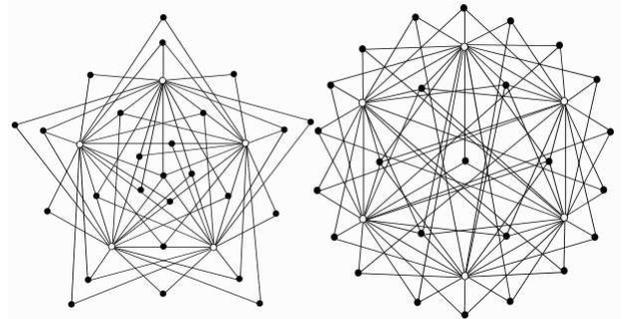} \caption{$d\geq 3$ graphs beyond the MSC. The left graph corresponds to the $|0_L\rangle$ state of the 31 qubit code with transversal $\exp{\left(-i \frac{\pi}{16} Z_L \right)}$ gate. And the right one is a graph corresponds to $|+_L\rangle$. In each graph all the black vertices are minimal elements which are just generators of the corresponding $\mathcal{M}$, and all the white vertices are not in $\mathcal{M}$.} \label{31qubit} \end{figure}

Note that the graphs corresponding to $|0_L\rangle$s of the code
always have distance $3$ for any $m$, and the graphs corresponding
to $|+_L\rangle$s of the code always have distance $4$ for any $m$.
It is straightforward to show that for any $m$, only $Z$ appears on
all the qubits in $\mathcal{M}$ for both $|0_L\rangle$ and
$|+_L\rangle$. This then gives two series of $\delta>2$ graphs
beyond the MSC. The graphs for $m=4,5$ are shown in
Fig.~\ref{15qubit} and \ref{31qubit}.

\subsection{$LU\Leftrightarrow LC$ property for $\delta>2$ graphs}

It is natural to ask whether we could use the same strategy to prove
$LU\Leftrightarrow LC$ for those $\delta>2$ graph states beyond the
MSC as we did for $\delta=2$ graphs.

First of all, it is noted that a similar deletion of a degree $d-1$
vertex is possible. Take the above $\delta=3$ graph in
Fig.~\ref{distancegeq3} A6 for instance. Denote the two white
vertices by $2,3$, and the degree two vertex which connects to $2,3$
by $1$. Then the stabilizer of $1,2,3$, up to LC, can be written as
\begin{equation} Z_1Z_2Z_3,X_1X_2R_j,X_1X_2R_k \end{equation} where
$R_j,R_k$ denotes the operators on the other qubits apart from
$1,2,3$.

Now recall the $n$-qubit quantum code ${Q}_e^{(n)}$ with stabilizer
$\mathcal{S}({Q_e^{(n)}})=\{I^{\otimes{n}},Z^{\otimes{n}}\}$ is a
quantum version of the $[n,n-1,2]$ classical binary zero-sum code
(or even weight code). The basis of ${Q}_e^{(n)}$ can be simply
chosen as all the codewords with even weight, and any of the $n$
qubits can be regarded as a parity qubit of the other $n-1$ qubits.
In this sense, $Q_e^{(n)}$ encoding $n$ qubits into $n-1$ qubit, we
will always choose the basis for $n-1$ logical qubits to be that of
omitting the first qubit. For instance, if $n=3$ (as mentioned in
graph A6 of Fig.~\ref{distancegeq3}, the stabilized subspace of
$Z_1Z_2Z_3$ is spanned by \begin{equation}
\{|0_10_20_3\rangle,|1_10_21_3\rangle,|1_11_20_3\rangle,|0_11_21_3\rangle\},
\end{equation} which could be viewed as two logical qubits:
\begin{eqnarray}
\{|00\rangle_L=|0_10_20_3\rangle,|01\rangle_L=|1_10_21_3\rangle,\nonumber\\
|10\rangle_L=|1_11_20_3\rangle,|11\rangle_L=|0_11_21_3\rangle\},
\end{eqnarray} where the first physical qubit acts as a parity qubit
of the other two.

Any LU operation $\mathcal{F}_n=\bigotimes_{i=1}^{n} F_i$ where each
$F_i$ is diagonal preserves $Q_e$ and will induce an diagonal
logical operation $F_L$ on the $n-1$ logical qubits.

[\textbf{Proposition 2}]: For an $n$-qubit even weight code $Q_e$,
if $F_L\in \mathcal{L}_{n-1}$, then $F_i\in\mathcal{L}_1$ for all
$i=1,\ldots,n$.

[\textbf{Proof}]: Since $F_L$ is diagonal, it preserves $Z_i$ for
all $i=2,\ldots,n$. Let $F_i=diag\{1,e^{i\theta_i}\}$, direct
calculation shows $F_LX_2F_L^{\dagger}\in \mathcal{G}_{n-1}$ if and
only if both $e^{2i\theta_1}=\pm 1$ and $e^{2i\theta_2}=\pm 1$, i.e.
$F_1,F_2\in\mathcal{L}_{1}$. Similar procedure works for
$i=3,\ldots,n$.$\square$

However, generally $F_L$ is a non-local operation on the $n-1$
logical qubits, contrary to the $\delta=2$ case, where the local
operation can only induce a local operation on the single logical
qubit. Therefore, it is non-trivial to delete a degree $d-1$ vertex.

A possible way to fix this problem may be to further investigate the
effect of some non-local gates (in this example, two-qubit gates)
which relate the two graph states. Then we could use
\textbf{Proposition 2} to prove $LU\Leftrightarrow LC$ for the
original graph before deletion of the vertex. This idea does work in
the case of the particular structure of the graph A6 in
Fig.~\ref{distancegeq3}, after a subtle analysis on the structure of
$\mathcal{S}$.

Our \textbf{Proposition 2} takes the first step to investigate the
$LU\Leftrightarrow LC$ property for $\delta>2$ graphs beyond the
MSC, which is also based on the subgraph structure. However, it is
not our hope that the idea of induction will final lead to a
solution to the most general case. For instance, it is noted that
$|{\psi}\rangle$ satisfying the MSC does not necessarily mean
$\mathcal{S}(|{\psi}\rangle)=\mathcal{M}(|{\psi}\rangle)$ , although
exceptions are likely rare. We have found only two LU inequivalent
examples for $n\leq 9$, which are shown below in Fig.~\ref{sneqm}.

\begin{figure}[htbp]
\includegraphics[width=2.30in]{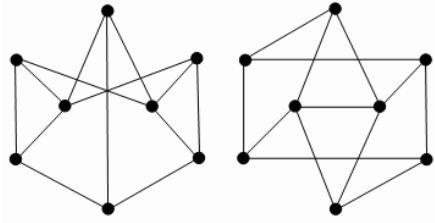}
\caption{Two $n=8$ graphs satisfying the MSC, but
$\mathcal{S}(|{\psi}\rangle)\neq\mathcal{M}(|{\psi}\rangle)$.}
\label{sneqm}
\end{figure}

Note both of the two graphs in Fig.~\ref{sneqm} are of $n=8$. There
exist two graphs satisfying the MSC but $\mathcal{S}\neq\mathcal{M}$
for $n=8$, however there does not exist any graph of this property
for $n=9$. This interesting phenomenon implies that the structure of
$\mathcal{M}$ is a global rather than a local property of graph
states, which cannot be simply characterized by the idea of
induction.

\section{Conclusion and discussion} \label{sec:concl}

In this paper, we broaden the understanding of what graph and
stabilizer states are equivalent under local Clifford operations. We
prove that $LU\Leftrightarrow LC$ equivalence holds for all graph
states for which the corresponding graph contains neither cycles of
length $3$ nor $4$. We also show that $LU\Leftrightarrow LC$
equivalence holds for distance $\delta=2$ graph states if their
corresponding graph satisfies the MSC after deleting all the degree
one vertices. The relation between our results and those of Van den
Nest et al.'s is summarized in Fig.~\ref{paperdiag}. It is clearly
seen from the figure that graphs in area $D$ have no intersection
with those in $C$, i.e. graph states of distance $\delta=2$ are
beyond Van den Nest et al.'s MSC. The intersection of graphs in area
$B$ and $C$ are graphs without degree one vertices as well as cycles
of length $3$ and $4$.

We find a total of $58$ $\delta>2$ graphs beyond the MSC up to
$n=11$, via numerical search; among these, only $10$ are of
$\delta=4$ while the other $48$ have distance $\delta=3$. This
implies that $\delta>2$ graphs beyond the MSC are rare among all the
graph states, and are not easy to find and characterize. However, we
also explicitly construct two series of $\delta>2$ graphs using
quantum error correcting codes which have non-Clifford transversal
gates. We expect that the existence of other such quantum codes will
provide insight in seeking additional $\delta>2$ graphs beyond the
MSC. All graph states discussed in this paragraph belong in area $E$
in Fig.~\ref{paperdiag}. For most of the graphs in area $E$, the
$LU\Leftrightarrow LC$ equivalence question remains open. We
discussed some possibilities for resolving this equivalence question
in Sec. IV, using even weight codes rather than the simple
repetition codes.

Our main new technical tool for understanding $LU\Leftrightarrow LC$
equivalence is the idea, introduced in Sec.~\ref{sec:maintheorem},
of encoding and decoding of repetition codes. We hope that this
tool, and our other results, will help shed light on the unusual
equivalences of multipartite entangled states represented by
stabilizers and graphs, and the intricate relationship between
entanglement and quantum error correction codes which allow
non-Clifford transversal gates.


\end{document}